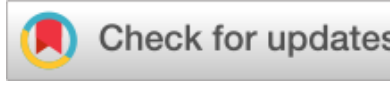

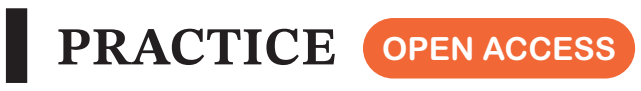

PRACTICE OPEN ACCESS

# Introducing Feedback Thinking and System Dynamics Modeling in Economics Education

Oleg V. Pavlov[1] | Robert Y. Cavana[2] | I. David Wheat[3] | Khalid Saeed[1] | Michael J. Radzicki[1] | Brian C. Dangerfield[4]

[1]Worcester Polytechnic Institute, Worcester, Massachusetts, USA | [2]Wellington School of Business and Government, Victoria University of Wellington, Wellington, New Zealand | [3]University of Bergen, Bergen, Norway | [4]Loughborough Business School, Loughborough University, Loughborough, UK

**Correspondence:** Robert Y. Cavana (bob.cavana@vuw.ac.nz)

## ABSTRACT

System dynamics is a methodology that is widely used in many academic fields. It explains the behavior of social and economic systems with models that capture complex causality and feedback effects. This paper discusses the opportunities and barriers for introducing feedback thinking and system dynamics models in the economics curriculum. We start by providing a pricing feedback model that illustrates some of the benefits that system dynamics can provide in enhancing economics education. Then we summarize the experiences of each of the authors in teaching system dynamics in economics educational programs. This includes different approaches to teaching economics with system dynamics that depend on the learning objectives, the preparation of students, and the background of the instructor. We also develop a four-level course hierarchy for using system dynamics in economics teaching. We then point out the tradeoffs that instructors must consider as they introduce new pedagogies for delivering economics material. Finally, we provide concluding comments with some suggestions for future work. The expected audiences for this paper are instructors as well as graduate students who are considering academia as a profession.

## 1 | Introduction

System dynamics, pioneered by Professor Jay Forrester at MIT in the late 1950s (Forrester 1961), has evolved into a research discipline with its own academic society, a peer-reviewed journal, degree-granting educational programs, and a collection of well-regarded textbooks. System dynamics has been widely applied across various fields of research and instruction including management, natural sciences, sustainability, and public policy. While there exists a substantial body of literature that applies system dynamics to economics (e.g., Cavana et al. 2021; Fiddaman 2002; Forrester 1973b; Harvey 2002; Low 1980; Mass 1975; Meadows 1970; Meadows et al. 1972; Pavlov et al. 2005; Radzicki 2020; Saeed 2020; Scrieciu et al. 2021; Sterman 1986, 2000; Uehara et al. 2016; Wheat 2009; Yamaguchi and Yamaguchi 2021), system dynamics is not part of the standard economics curriculum. This raises questions regarding the discrepancy between the versatility of system dynamics and its underrepresentation in economics education. Despite the recognition of “feedback systems” by economists dating back to the 18th-century (see Mayr 1971; Richardson 1991) and compelling calls for “rethinking macroeconomics” following the 2008 crash (e.g., Sachs 2009; Stiglitz 2018), the integration of system dynamics into economics teaching has been hindered by a lack of awareness among economists about its principles and educational affordances.

In general, very little guidance currently exists regarding how system dynamics may be adapted to economics. In Harvey's (2021) review of our co-edited *Feedback Economics* book (Cavana et al. 2021), he goes on to say (p. 363) that “… this is no longer the case as the book is an incredible resource for anyone who believes that real-world economic phenomena can be properly understood by interdependencies, feedback loops, and endogenous change. This is the theme throughout and there are instructions, applications, examples, and beautifully illustrated diagrams in every chapter." Sixty years later, this book certainly lends support to Professor Harvey Wagner's insightful comments in his review in *Management Science* of Forrester's (1961) *Industrial Dynamics* book that "...Forrester could very well be credited with providing the missing link between the grand conceptions of classical Continental economists and the imaginative inventions of modern electronic wizards" (Wagner 1963, 184).

Our *Feedback Economics* book is a good starting point for building bridges between the system dynamics and economics scholarly communities. It is not the only teaching resource available, but it goes a long way to complement standalone system dynamics courses in economics departments as well as a complementary resource for existing specialized economics courses whose instructors want to teach their material with a "feedback and simulation" lens characterized by the system dynamics approach. In this paper, we hope to demonstrate the value of using system dynamics in teaching economics based on our collective experiences and provide a guide to the resources available to do this.

Additionally, this paper asserts that system dynamics can be successfully integrated with economics instruction, often complementing rather than replacing traditional economics pedagogy. Causal diagrams and system dynamics simulations can enrich traditional comparative statics analysis by allowing students to explore the transitional paths between successive equilibrium points within an economic system (an illustration of this is provided in the next section). While very few economists are familiar with system dynamics, they already employ practices that are foundational to system dynamics such as project-based learning (e.g., Ghosh 2013) and numerical simulations (e.g., Gorry and Gilbert 2015). We are confident that as economists continue searching for innovative pedagogical approaches that enhance the accessibility and relevance of economics material, feedback thinking and system dynamics modeling will enter the mainstream economics education thus fostering a deeper understanding of complex economic systems by students and better preparing them for real-world challenges.

This paper also contributes to the literature on the pedagogies of system dynamics (e.g., Bosch and Cavana 2018; Richardson 2014; Schaffernicht and Groesser 2016) and economics (e.g., Allgood et al. 2015; Asarta et al. 2017; de Muijnck et al. 2021; Frank 2002). In the following section, we provide an illustration of a pricing feedback module. Then we discuss the state of the art in teaching economics with system dynamics by drawing on the authors' experiences and published literature where relevant. In the pedagogy integration section, we develop a course framework that aligns closely with the system dynamics competency model of Schaffernicht and Groesser (2016). The framework suggests practical instructional strategies for introducing feedback thinking and system dynamics modeling into economics education. We also briefly explain the barriers to introducing new instructional methods in economics. Some suggestions for future work are outlined in the concluding section.

## 2 | Teaching Economics With System Dynamics—An Illustration

Economic analysis with system dynamics, or *feedback economics* (Cavana et al. 2021), explains the behavior of economic systems by examining their structure. The approach uses reference modes, causal loop diagrams, and numerical simulations of stock-flow consistent models. There are several ways for introducing feedback thinking and system dynamics modeling in economics education, as we discuss further in this paper. As an illustration, we start with an instructional module on price and quantity adjustments.

This module extends conventional comparative statics analysis that uses a supply-and-demand graph, such as in Figure 1, commonly found in introductory economics textbooks like Mankiw (2021). In the graph, the initial equilibrium is ($Q_1$, $P_1$). When the demand curve shifts to the right due to some external factors, the equilibrium quantity and the equilibrium price increase to new values ($Q_2$, $P_2$).

Figure 2 shows a corresponding feedback explanation of price adjustment that has been discussed in several publications (e.g., Jägerskog 2021a, 2021b; Joffe 2021; Lyneis 2003; Wheat 2007; Whelan and Msefer 1996; Yamaguchi 2022). The graph, called a *causal loop diagram*, follows diagramming conventions of system dynamics (Lane 2000). It shows causal relationships between key variables. A positive arrow means that the cause and effect move in the same direction. When the cause and effect move in opposite directions, a negative arrow is used. A delay between the cause and effect is shown as two dashes perpendicular to an arrow. Circular causality forms reinforcing and balancing *feedback loops*, also referred to as positive and negative loops. In Figure 2, there are two balancing loops, B1 and B2, that control price adjustments. Following Joffe (2021), we can trace the effect of a positive demand shock. When product demand increases, the variable "shift in demand" goes up, leading to an increase in the quantity demanded. More competition between buyers for the product causes its price to increase. A higher price will entice more sellers to supply this product to the market, which is shown as a positive arrow between price and "quantity supplied." As more suppliers compete in the market, the price drops (a negative arrow on the left). As the price decreases, the quantity demanded increases (a negative arrow on the right). Following this mechanism, the price may adjust several times before it settles to a new equilibrium.

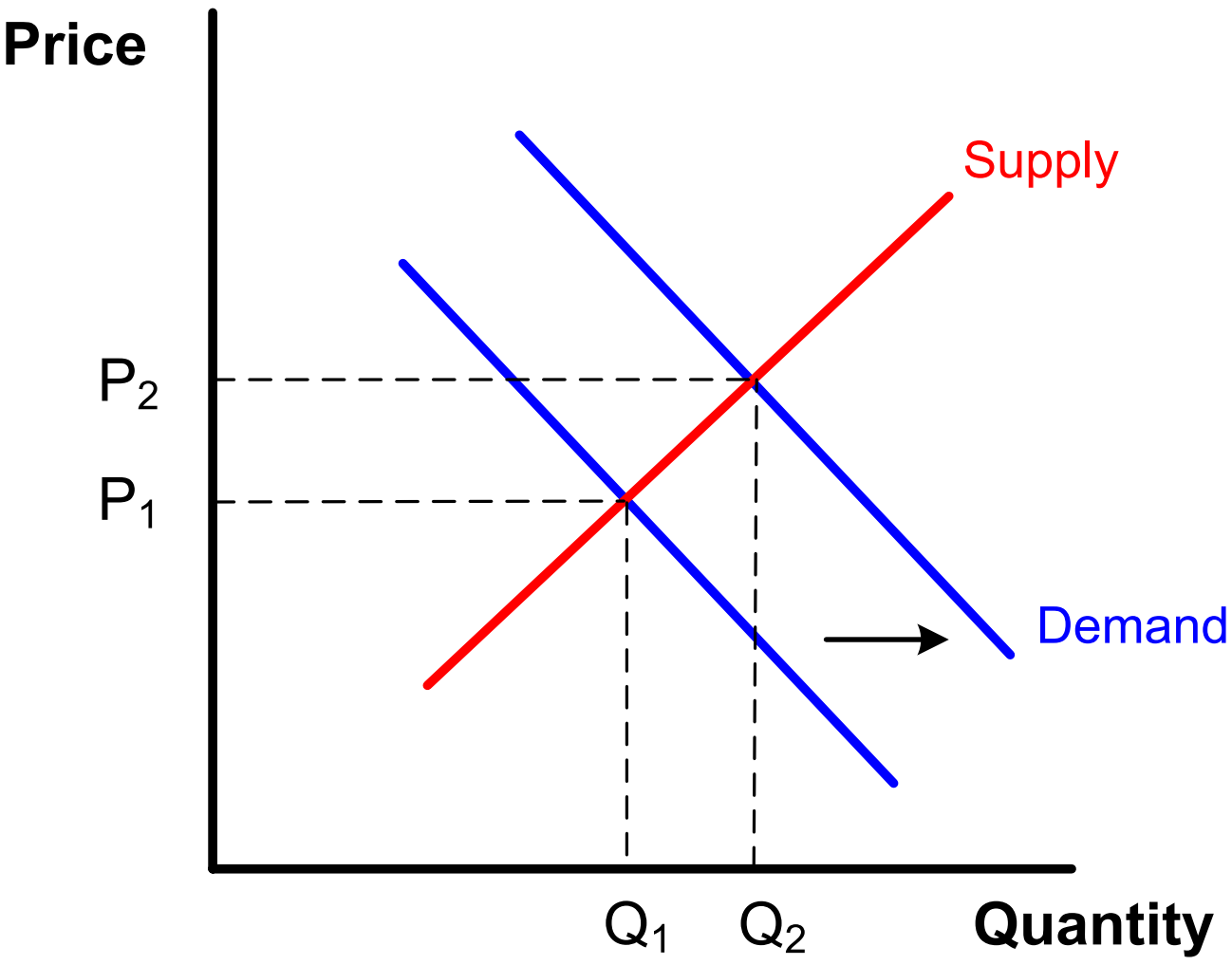

**FIGURE 1** | The comparative statics analysis using a supply-and-demand graph that explains how an increase in demand raises the equilibrium quantity from $Q_1$ to $Q_2$ and the equilibrium price from $P_1$ to $P_2$.

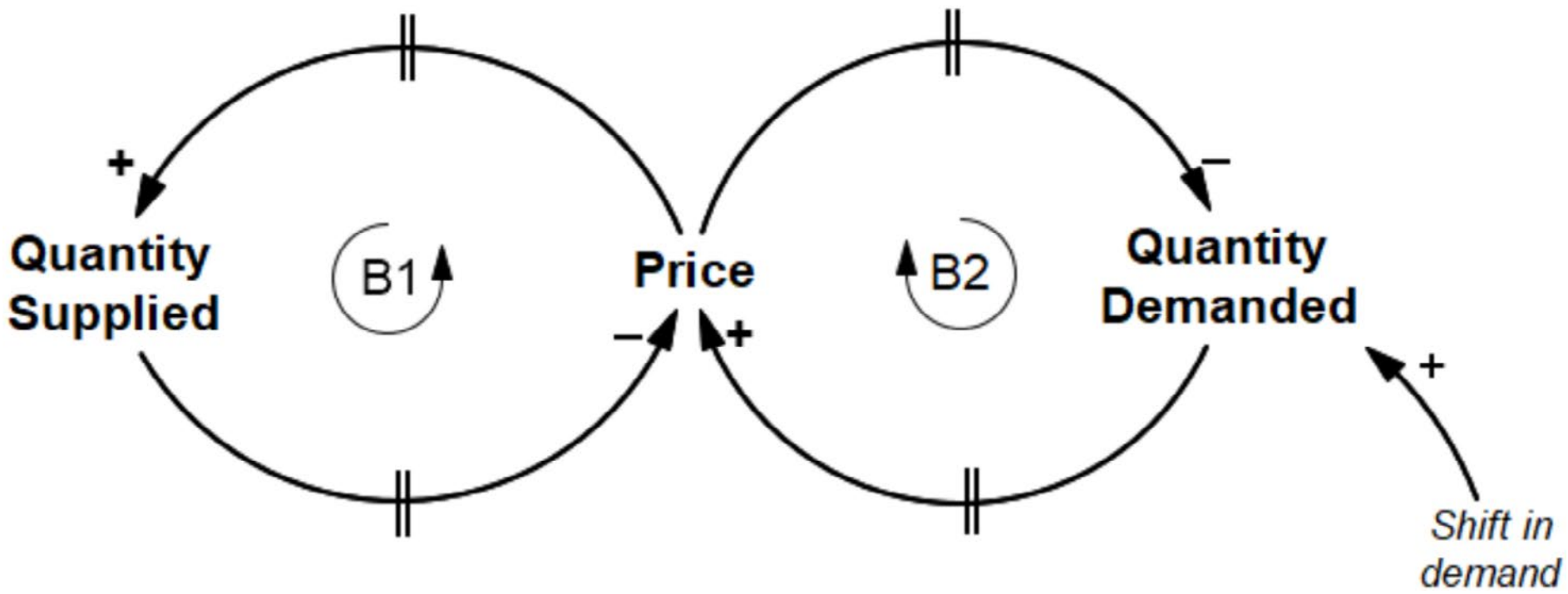

**FIGURE 2** | A simplified diagram that shows the causality between supply, demand, and price. The circular causality forms two balancing feedback loops, B1 and B2.

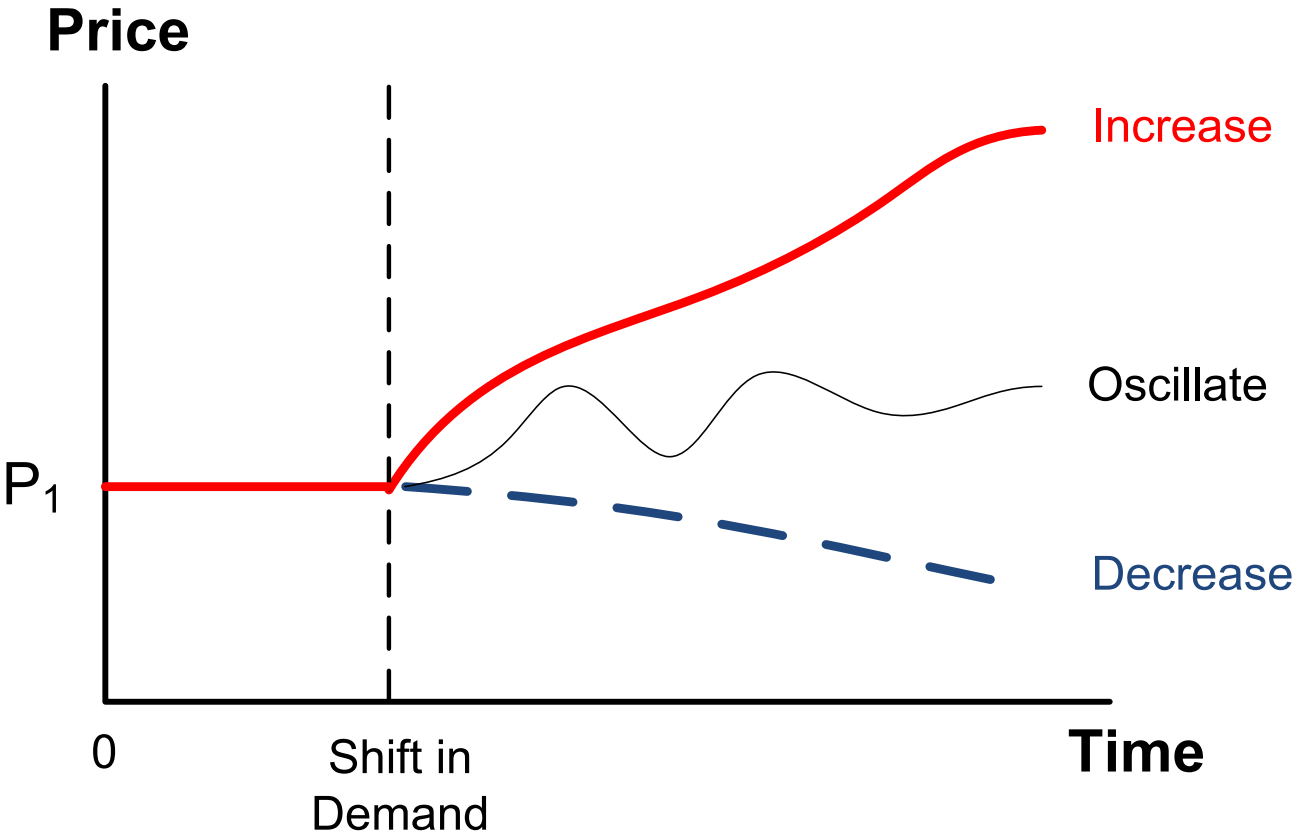

**FIGURE 3** | A reference mode graph for price behavior.

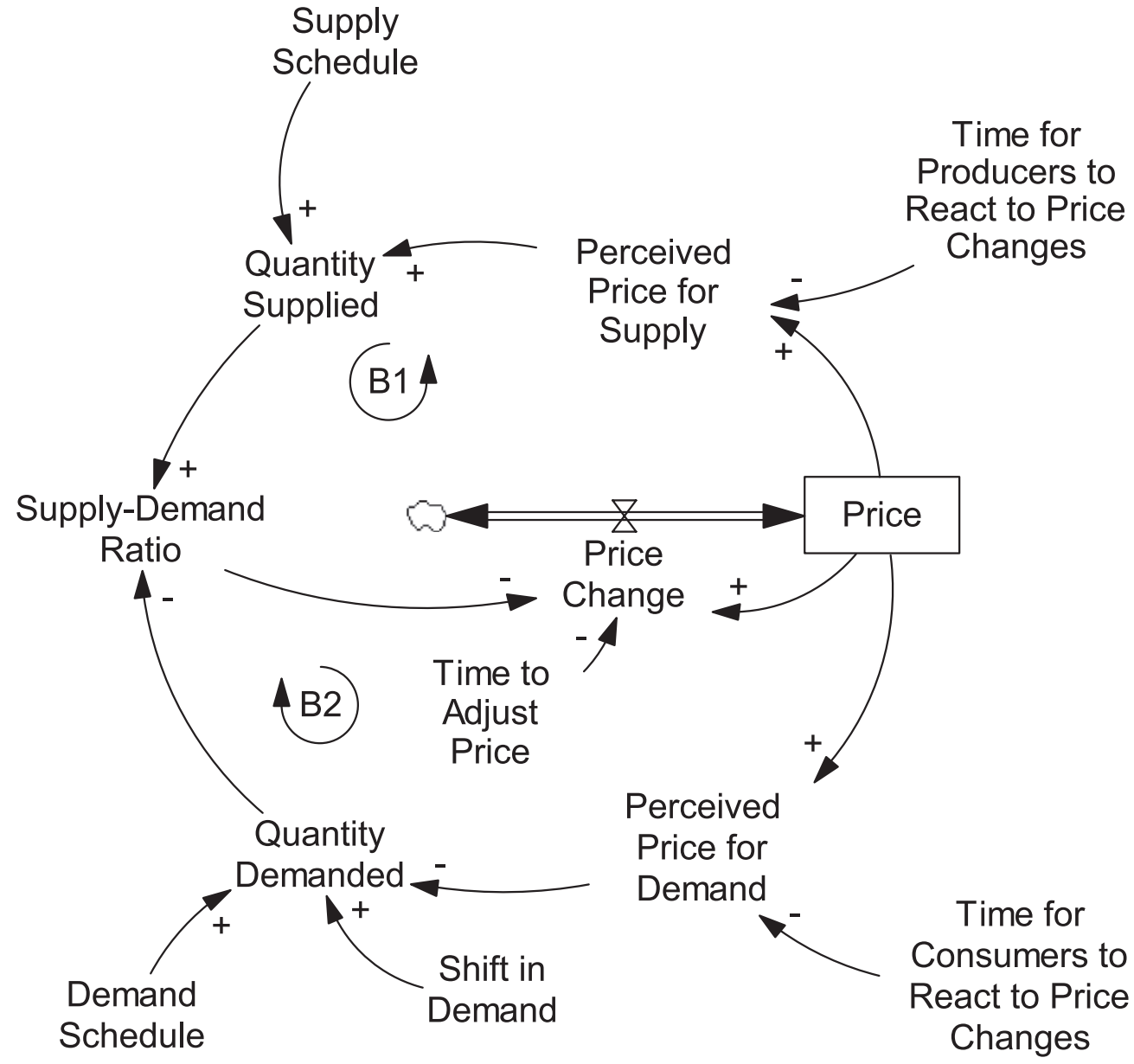

**FIGURE 4** | The stock and flow structure of the supply-and-demand model. Adapted from Lyneis (2003).

To guide a classroom discussion, the instructor may also use a graph called a *reference mode*, or a *behavior-over-time graph*, which is commonly used in feedback analysis to explore out-of-equilibrium behaviors. As students consider price behavior, they may start by drawing the price fixed at $P_1$ until the demand shock is introduced into the system, see Figure 3. Then, students may hypothesize that in response to the shift in demand the price can either increase, decrease or oscillate. To test these hypotheses, students can run numerical simulations.

For the simulations, the feedback structure in Figure 2 must be implemented as a computational model. Figure 4 shows *the stock and flow diagram* for a *system dynamics model* adapted from Lyneis (2003). The model includes two balancing loops, which correspond to the feedback structure in Figure 2. The variable "Price" is shown as a rectangle, meaning it is a stock, or a state variable. Because system dynamics models are continuous-time models, price changes according to an integral equation:

$$\text{Price} = \int (\text{Price Change})dt + P_1$$

where, $P_1$ is the initial equilibrium price from Figure 1. The flow variable "Price Change" is a function of the supply and demand imbalance; it is drawn as a bi-directional pipe connected to the stock. Model equations can be found in the Appendix A, and the model file is included as a supplementary attachment. Graphical simulation results are shown in Figure 5. The simulation starts in equilibrium when price $P_1 = 25$ \$/unit and quantity demanded equals quantity supplied at $Q_1 = 50$ unit/day. At time 10, demand shifts up by 10 unit/day. After some adjustments, the market settles in a new equilibrium, $P_2 > P_1$ and $Q_2 > Q_1$, as also predicted by the comparative statics analysis in Figure 1. At this point, students can observe that the price trajectory in Figure 5 corresponds to the oscillating scenario in Figure 3.

While this module explains a relatively simple case of price adjustments in response to a demand shift, system dynamics models can explain significantly more complex situations. Additional examples can be found in an edited volume on *Feedback Economics* by Cavana et al. (2021).

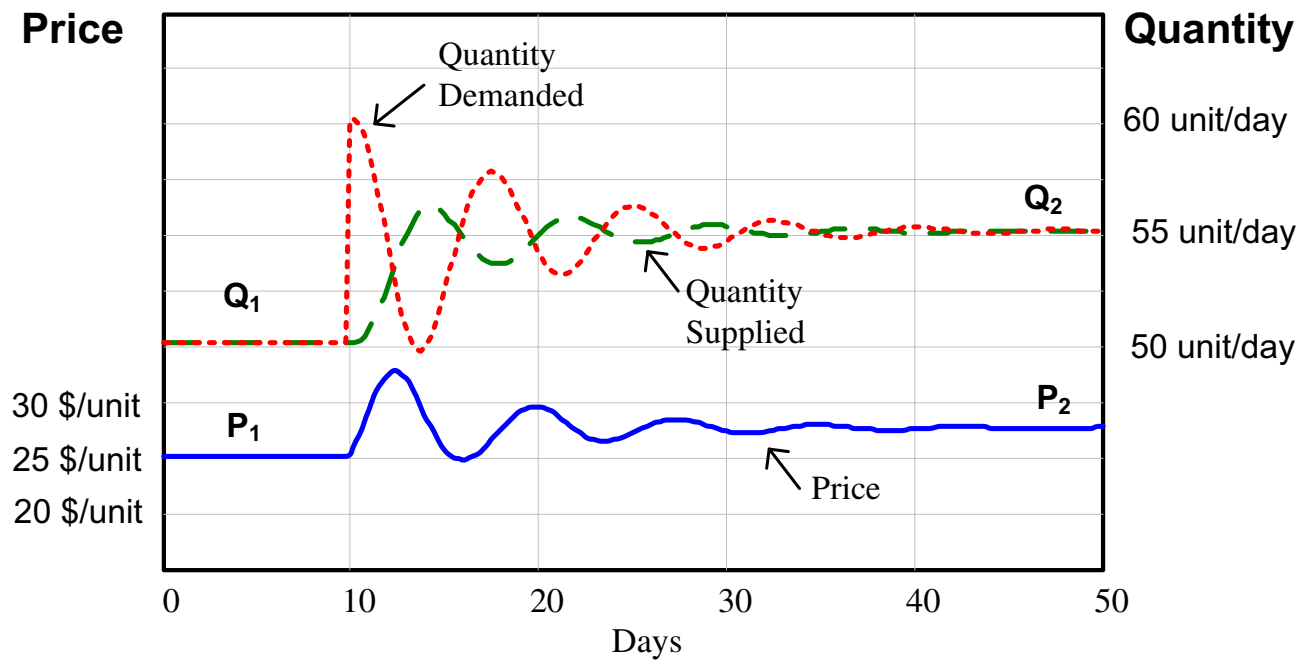

**FIGURE 5** | The output of the supply and demand model in Figure 4. The simulation starts in equilibrium $(Q_1, P_1)$ and then settles to a new equilibrium $(Q_2, P_2)$ after experiencing a demand shock at time 10.

## 3 | Authors' Teaching Experiences and Lessons Learned

At the 2022 International System Dynamics Conference, hosted by the Frankfurt School of Finance and Management, we convened a roundtable titled "Teaching Economics with System Dynamics." This section is based on an edited transcript of the discussion. Although Brian C. Dangerfield could not attend the roundtable, he contributed to this paper as a co-author. Drawing on over 200 years of our combined teaching experience across various institutions and countries, this section offers a broad perspective on integrating system dynamics into economics education.

### 3.1 | Oleg V. Pavlov

Jim Lyneis, my former colleague at Worcester Polytechnic Institute (WPI), taught principles of microeconomics by adding elements of system dynamics to a traditional economics course, a strategy he called a *merged approach*. As he described it in an article (2003), he thought that there were three motivating factors for this strategy. First, some topics are best taught in a traditional way, rather than with system dynamics. Second, many economics concepts create the foundation for system dynamics models, and therefore should not be left out. And third, he thought that as some students would continue their economics education, they needed to know standard economics methods and models. Because it was a large class, it did not include hands-on experiences with model development and simulations.

Building on Lyneis' experience, in my teaching, I often follow the standard economics outline and use mainstream economics textbooks, but I add causal loop diagrams and self-contained system dynamics modules. For example, during the COVID-19 pandemic, I adapted the "epidemic game" originally developed by William Glass-Husain (Fisher 2007; Glass 1991) as a system dynamics module for an online microeconomics course. Students played the game and gathered synthetic data during synchronous online sessions and then built simple epidemic models that they calibrated to the collected data. This activity was followed by a discussion of such traditional economics topics as externalities and the cost–benefit analysis of public goods, including vaccination programs. In addition to reading a microeconomics textbook, students experimented with their simulation models, trying to "contain" the "pandemic." On another occasion, I used a module on fisheries for discussing the tragedy of the commons. The activity involved building a relatively simple system dynamics model. But I later concluded that building even such a simple model turned out to be too complex as an activity for a large introductory course. Whenever I use system dynamics modules, students work in groups, which they enjoy.

As anyone who teaches at WPI, I also supervise student projects that can be as short as 7 weeks or as long as an entire academic year. These projects, which are often completed in groups, allow students to apply system dynamics methods jointly with economic theory to various problems. Depending on the students' backgrounds, majors, and interests, projects utilize qualitative or quantitative system dynamics. Some projects may lead to publications (e.g., Katsamakas et al. 2024; Pavlov et al. 2015; Pavlov and Sardell 2021).

### 3.2 | Robert Y. Cavana

Although prior to joining academia I had been employed as a professional economist, my system dynamics teaching has mostly been undertaken in a School of Business and Government and occasionally in a School of Engineering. I have been teaching at undergraduate, graduate and executive development levels for over 30 years. Typically, I taught courses which covered qualitative and quantitative methods of system dynamics using introductory material mostly from Maani and Cavana (2007). This included some or all of the following: problem structuring; causal loop modeling; group model building; stock and flow diagramming; intro to Vensim or Stella simulation; management, economics and sustainability examples; analyzing dynamic models; technical aspects of simulation modeling; model validation and sensitivity analysis; policy analysis and modeling; strategy development and scenario analysis.

The courses could involve combinations of lectures, tutorials, computer laboratories, simulation games, assignments and group presentations. Typically, the group projects included a case on a current media topic at undergraduate level and a selected policy-oriented topic for the group projects at the graduate level (generally with an economics lens or focus). Examples of the student cases that expanded into publications include the "alcohol in New Zealand supermarkets case" outlined in Mabin and Cavana (2024), based on an introductory "systems thinking and decision making course" at undergraduate level; as well as strategic modeling projects undertaken with MBA students related to the analysis of "farm woodlot options" and the design of a "policy making framework for the New Zealand wine industry" (Cavana et al. 1996, 1997).

Lessons learned from these teaching experiences are reflected in some of the comments provided by participants that these courses provide:

- an enjoyable and exciting introduction to powerful techniques for modeling complex behaviors, in organizations and their environment;
- an intuitive way of thinking that allowed application of a simple modeling tool to provide excellent results;
- its easy to think you are a systems thinker, it is not until you get into a model, and start to realize the significance of manipulating systems variables, that you realize both the power and skill involved in systems thinking skills to address real problems.

Overall I have concluded that teaching systems thinking and system dynamics to students training to be an economist or similar business professional, can complement education in traditional econometric and statistical methods by: focusing on viewing problems from a systems and feedback thinking perspective; learning and applying systems modeling tools and methods; assistance with identifying problems and behavior over time patterns; identifying and engaging with stakeholders affected or involved; boundary selection for soft or hard variables; working in groups; validation and analysis of dynamic problems; and presenting and implementing policy recommendations.

### 3.3 | I. David Wheat

In the late 1990s, I was invited to be an adjunct instructor at a community college in the United States. After a few semesters watching my students struggle with a standard macroeconomics textbook and its collection of seemingly disconnected static models, I decided to take system dynamics into the macro classroom.

I still teach that course but now use my own textbook that features a full SD-based macro model (Wheat 2025). The model is the students' flight simulator for the U.S. economy. The course covers traditional macro topics but presents them in a way that reveals the economy as a dynamic feedback system. Students use the free *Stella Online* software to build simplified versions of sub-models, and they engage in "what if …" experiments with the online simulator. The most important feedback loop in the whole course is one that ties together the supply side and the demand side of the economy. Most introductory macro textbooks describe demand-side business cycles and supply-side growth as if they were two planets orbiting in separate solar systems. We use the same model for both cycles and growth and emphasize that the long run is just a succession of short runs. I also show students the implicit negative feedback thinking in economic policies. An endogenous policy attempts to redirect economic performance towards a goal.

In addition, I teach SD-based economics courses to graduate students in Europe. In Norway, the course is focused on policy design. Students adapt the textbook flight simulator model to other countries and experiment with alternative policies. The course in Ukraine emphasizes economic dynamics, using small macro and micro models. For many years, I taught a monetary policy course for Lithuanian students in a financial economics program. Particularly useful was a *monetary policy game* developed for the textbook simulator model.

Similar instructional methods can also be used in workshops tailored for particular groups. In a workshop for policymakers, we converted a regional SD-based model to a national model. The participants lacked prior system dynamics training and got just enough during the workshop to develop a feel for the model's dynamics. At the other extreme, my hands-on workshop at system dynamics conferences is for participants with a system dynamics background who want to know about modeling a macroeconomic system.

For more information about these courses, the readers can see my articles (Wheat 2007, 2010, 2017, 2023; Wheat et al. 2021) and the Supporting Information for my new textbook (Wheat 2025).

### 3.4 | Khalid Saeed

My graduate work at MIT coincided with the time the system dynamics group there was working on the National Modeling project that attempted to reinvent macroeconomics by building its models around managerial roles and explaining macro-behavior from micro-structure (Forrester 1989). In my Ph.D. work I applied system dynamics to economic development problems. Working simultaneously in economics and system dynamics, I think of them as two names of the same problem domain (Amissah et al. 2020).

After graduating from MIT, my teaching focused on graduate system dynamics courses at the Asian Institute of Technology and then graduate and undergraduate system dynamics courses at Worcester Polytechnic Institute, WPI. My students built models of cases and news-stories that articulated economic issues. I started teaching undergraduate development economics and environmental economics at WPI in 2004 and, like Lyneis (2003), I felt compelled to cover traditional topics, but I also included learning from my own system dynamics models as an alternative approach and encouraged students to critically evaluate all perspectives. In the development economics course, we discussed poverty alleviation (Saeed 2020), governance (Saeed 1990), and political instability (Saeed 1986) based on my research as well as Forrester's Urban Dynamics (Saeed 2015). The environmental economics course covered World Dynamics (Forrester 1973a), resource use (Saeed 2013), and environmental restoration (Saeed 2004). I often use games such as Fishbanks in the environmental economics course and also Global Warming, People Express and Beer Game in an undergraduate course titled Games for Understanding Complexity.

In my chapter (Saeed 2021) in *Feedback Economics*, I describe system dynamics models of the classical growth theory. I have used these models both in my courses on economics and system dynamics. In general, the models that I use in my teaching are simple and parsimonious. They help me combine economics and system dynamics instruction, explain economics concepts to students, and add system dynamics to teaching economics. In my experience, system dynamics models of economic theories can liven up a class by bringing active learning into it. I also utilize extensively the storytelling feature of the modeling software Stella Architect from *isee systems* that allows a step-by-step presentation of a model.

I have recently started using learning environments (see, e.g., Qudrat-Ullah 2019, 2020; Tadesse et al. 2021) for directed assignments designed for students unfamiliar with system dynamics. I have a new course on ecological economics and all my homework is in the form of learning environments, which are published on the *isee systems* website. Students are coached to acquire reading literacy for recognizing feedback loops and stock flow relationships in a system dynamics model to be able to use the learning environments. They are not required to build models but are directed to set up experiments with the supplied models using their respective user interfaces, and answer questions interpreting the simulations arising from those experiments.

### 3.5 | Michael J. Radzicki

It seems to me that there are different ways to tie system dynamics to economics. You can build models from the written descriptive theories that are quite common in economics. I think it is very powerful that we can translate into dynamic models something like the work of Thorstein Veblen and John Kenneth Galbraith by explaining what they said in terms of feedback loops.

Static theories of course have been traditionally used in economics. Economists developed the static machinery of comparative statics analysis that allows us to go from this equilibrium to that equilibrium, but you never see what happens in between. But of course, that is where all the interesting stuff happens. So, I like with the undergraduates to take a traditional model and then recast it in a dynamic perspective and then reproduce the static results, and then look at what happens in between the equilibria. And sometimes it is quite uninteresting, you just get a smooth transition from one to the next, but a lot of times that is because the stock and flow structures are inaccurate. And then you start to say, well, what if we fix the stock and flow structure, what happens now? Now you start to see interesting things.

Take macroeconomic analysis of Keynes, whose big motivation for doing macroeconomics was the fallacy of composition. We have macroeconomics because you cannot attribute the behavior of the whole as being the same as the behavior of any of the parts. That is a logical fallacy, right? A classic example is if I save more, that is good. But if everybody saves more, that is bad because you throw the economy into a recession. So, when you do the paradox of thrift in a dynamic model, you see that when people are thrifty, they save more because you got the stock-and-flow structure. And you can show these kinds of stunning results such as if you double your thriftiness, you save the same amount because you save a larger fraction of the smaller GDP. Rarely did I see in graduate growth theory courses a graph with time on the horizontal axis, even though they are doing dynamic models and continuous time. Therefore, recasting even difference equation models into system dynamics models is often insightful. With dynamic simulations you can really reveal a lot of interesting stuff.

Speaking of building original models from scratch, it is a whole different thing because it requires students to learn how to do system dynamics. It is the audience that determines as to what I am trying to accomplish and the approach I might take.

### 3.6 | Brian C. Dangerfield

My teaching experience has been solely in Management Science but that embraced a significant chunk of system dynamics. Although courses were rooted in the quantitative aspects of management, some economics students took them as electives. Thus, over the years, I could be facing between 25% and 40% of students who were economics majors. This was important because economics students usually had a stronger background in mathematics than did the management students.

In my final year Advanced Management Science course students were given 2–3 weeks to produce individually a working model based upon a provided brief. The task would account for 30% of the mark for the unit, the remainder being accounted for by a formal examination. The teaching embraced causal loop diagrams, stock-flow diagrams, equation writing and the use of the system dynamics software Vensim. A useful resource was my chapter 3 (a Primer on System Dynamics) in the Brailsford, Churilov, and Dangerfield book (Dangerfield 2014).

One of the assignments was a description of a hypothetical closed economy in which an ageing population was causing a hit on government finances due to a lower tax take and higher government disbursements. The kernel of the project was a reduction in government cash balances which was to be addressed by increasing the personal tax rate. All students were expected to become familiar with the difference between stocks and flows, balancing units in equations, the dependency ratio, the propensity to consume, the downward multiplier, and a mass-balance check for internal model verification. It was clear that the economics majors had more prior knowledge of the economic concepts, but even they had issues in thinking through appropriate units of measure and the difference between stocks and flows, which one would expect would be an important point to cover in economics modules.

When introducing causal loop diagrams, a newspaper article was used to surface a dire situation in a retail sector where reducing profits meant businesses were laying off employees leading to reduced household incomes and so further reduced spending on retail goods. This is a positive loop in decay mode and the economics students were faster to appreciate that the loop polarity was unchanged whether the variable "Employment" was used or "Unemployment."

A standout pedagogical issue, which cropped up regularly, was that of choosing appropriate units of measure. It is clear students do not think about how something might be measured. And it suggests that introductory Quantitative Methods courses should include a treatment of measurement units for typical variables found in business and economics. Whilst the economics majors, with their stronger mathematical background, were able to formulate rate and auxiliary equations more easily than the management students, they often equally needed a push in producing the units of measure for each component in the equation.

## 4 | Integration of System Dynamics and Economics Pedagogy

Economics courses teach students verbal and quantitative reasoning and a set of standard economics tools that draw on methodologies from disciplines such as mathematics, philosophy, history, and psychology (Allgood et al. 2015; Asarta et al. 2017). When economics is taught with system dynamics, in addition to the economics material, students must also learn system dynamics skills (Schaffernicht and Groesser 2016), including terminology mastery, dynamic reasoning, model analysis, project initialization, model creation, model validation, policy design, and policy evaluation. In a stand-alone system dynamics program, these skills can be acquired by taking courses that follow a canonical teaching sequence (Richardson 2014). As students improve their modeling skills, they ascend the stages of competency (Schaffernicht and Groesser 2016) from beginner to advanced beginner, then competent, proficient, expert, and, finally, master. Because learning system dynamics skills takes time away from the economics material, an instructor who aims to teach economics with system dynamics must carefully consider how many systems concepts to introduce in the course.

We suggest that courses that teach economics with system dynamics can be grouped into four broad categories that require different levels of system dynamics competence. These categories are:

- *Merged courses*: These are conventional economics courses that include elements of system dynamics, but without the expectation that students become proficient in modeling. This name was proposed by Lyneis (2003).
- *Simulation-based courses*: In simulation courses, students use provided models, often presented as online learning environments (Qudrat-Ullah 2019), to explore topics in economics. Students develop skills of running and interpreting existing models.
- *Modeling courses*: Students learn how to develop and use system dynamics models.
- *Project-based courses*: These courses teach students how to initialize a project, create a model, validate models, conduct policy evaluation, and present results.

These course categories can approximately match to the first four stages in the Schaffernicht and Groesser (2016) competency framework: from Level I for beginner through Level IV for proficient. As Schaffernicht and Groesser point out, training can take an individual to the level of proficiency, but individuals become experts and masters through additional practice. As Table 1 explains, instructors may utilize different pedagogical approaches depending on the learning objectives, the duration of the course, and student and instructor backgrounds. The table also provides course examples within each approach.

Merged courses are the least disruptive to traditional economics pedagogy. Therefore, an instructor who is new to system dynamics may start with a merged course by adding just a few elements of feedback thinking to a conventional economics course to help students understand complex topics such as price adjustments. A simulation-based course that relies on previously prepared models and online learning environments is the next step that would offer students a deeper level of dynamic understanding. As the economics instructor learns more about system dynamics and becomes comfortable with the modeling process, then modeling and project-based course formats become possible.

### 4.1 | General Integration Considerations

In a series of articles, Jägerskog and colleagues (Jägerskog et al. 2019; Jägerskog 2021a, 2021b) studied the impact of causal graphs on how students learn economics. They assessed students' understanding of price adjustments when it was explained with the standard supply-and-demand graph as in Figure 1 versus a causal loop diagram as in Figure 2. The studies revealed that the way the topic was explained significantly affected student comprehension of causality because different visual representations communicate distinct aspects of the phenomenon, and therefore lead to distinct learning outcomes. Feedback-based explanations fostered a deeper understanding of the pricing mechanism compared to the conventional supply-and-demand graph. The studies also revealed that instructors who were new to systems thinking struggled using the causal loop diagram.

The integration of system dynamics and economics in the classroom is likely to be facilitated by several shared attributes:

- *Importance of graphs*: Both economics and system dynamics emphasize the importance of graphs. Economics instructors believe that graphs make the material more intuitive and accessible (e.g., Hey 2005). Graphical representations are also prominent in system dynamics (Lane 2000; Sterman 2000).
- *The use of numerical simulations*: Economists are increasingly recognizing numerical simulations as a powerful tool for enhancing economic intuition and learning in the classroom (e.g., Gorry and Gilbert 2015; Herz and Merz 1998). System dynamics allows students to explore complex economic dynamics with simulations without the need for advanced mathematical skills.
- *Project-based teaching*: Projects that engage students in active learning are popular in economic instruction (Asarta et al. 2017) and they are also a key feature of the system dynamics method.

However, there are significant barriers to introducing alternative pedagogical methods in economics. The first barrier is the cost to faculty in terms of time and effort that are required for learning and implementing effectively new teaching techniques (Allgood et al. 2015). Second, large class sizes in principles of economics courses present logistical challenges that make it difficult to implement active learning pedagogies (e.g., Frank 2002; Denny 2014). Third, given the limited classroom time, difficult decisions must be made regarding the coverage of topics and

**TABLE 1** | Course hierarchy and pedagogical approaches that combine economics and system dynamics.

| **Level I: Merged courses** | |
|---|---|
| Approach | *A conventional principles of economics course that includes elements of system dynamics* |
| Description | The classroom time allocated to system dynamics is minimal. Causal loop diagrams help to explain complex causality in economics. Students develop a basic understanding of systems concepts and system dynamics terminology. SD competency: Beginner. |
| Examples | *Principles of economics (Lyneis, Pavlov)*: An undergraduate course that follows a conventional economics course outline, but lectures are supplemented with causal loop diagrams or simple simulations. It may introduce some material that is usually not covered in an introductory course (such as inventory control) as a self-contained feedback module. A traditional economics textbook is used that may be supplemented with articles. |
| **Level II: Simulation-based courses** | |
| Approach | *An economics course that uses a large system dynamics model* |
| Description | A course is built around simulations with an existing system dynamics model. Emphasis is on economics and only minimal classroom time is allocated to explaining system dynamics terminology and concepts. Students learn how to use models and perform dynamic analysis using them. SD competency: Advanced beginner. |
| Examples | *Principles of macroeconomics (Wheat)*: This course has used conventional undergraduate macroeconomics textbooks but now uses Wheat's textbook (Wheat 2025) and covers traditional macroeconomics topics. System dynamics is not taught in detail. Students build simple models based on demonstrations in lectures and the textbook, and they also reproduce the sub-models of MacroLab Lite. Students use the full model as a "flight simulator" to explore key feedback mechanisms. *Monetary policy (Wheat)*: A graduate-level course that uses a system dynamics model called the Monetary Policy Game. Students analyze the model and may calibrate the model with data from various countries. |
| Approach | *An economics course that uses small system dynamics models* |
| Description | This approach relies on small models that have been prepared before the course. Assignments ask students to perform simulations with provided models. Students may be asked to extend models. SD competency: Advanced beginner. |
| Examples | *Development economics (Saeed)*: An undergraduate course that uses small models presented in learning environments and with the storytelling feature of the modeling software. *Ecological economics (Saeed)*: An undergraduate course based on small system dynamics models presented as online learning environments. *Macroeconomics (Radzicki)*: This undergraduate course recasts macroeconomics models from the dynamic perspective. |
| **Level III: Modeling courses** | |
| Approach | *A modeling course on topics in economics* |
| Description | This approach emphasizes building small system dynamics models. Students develop introductory knowledge of system dynamics modeling and software. Students may or may not have an economics background. SD competency: Competent. |
| Examples | *Model-based economic policy design (Wheat)*: A graduate-level course that teaches macroeconomic modeling to students with prior system dynamics training. Economic modeling provides an example of a system dynamics application. *Macroeconomics (Saeed)*: A graduate system dynamics course that uses system dynamics models of classical economic theories. The instructor prepared original models and lesson plans. *Macroeconomics (Radzicki)*: A graduate system dynamics course that is based on original economic texts. The course corrects the logic of post-Keynesian theories with simulation modeling. The audience are advanced graduate students in system dynamics who do not necessarily have an economics background. |
| Approach | *A modeling course on decision making, management, or sustainability studies* |
| Description | Students develop knowledge of system dynamics modeling and software. Project assignments are often completed in groups. SD competency: Competent. |

(Continues)

**TABLE 1** | (Continued)

| **Level III: Modeling courses** | |
|---|---|
| Examples | *Decision making and systems modeling (Cavana)*: Qualitative and quantitative system dynamics taught as part of courses within the business curriculum and engineering curriculum, at the undergraduate or graduate level. Topics included introductory system dynamics methods and group model building cases or projects on various topics. Audience: business, economics, and engineering students. *Management science (Dangerfield)*: Courses based on the quantitative aspects of management also cover causal loop diagrams, stock-flow diagrams, equation writing, units, and the use of system dynamics software. Students build system dynamics models. Audience: management and economics students. |
| **Level IV: Project-based courses** | |
| Approach | *Project-based teaching* |
| Description | Independent research projects that are tailored to students' interests and backgrounds. SD competency: Proficient. |
| Example | *Various projects (Pavlov)*: Students apply system dynamics methods jointly with economic theory to various problems. Usual duration: one to two semesters. Students work individually or in groups. The extent of modeling varies from causal loop diagrams to system dynamics models. |
| Approach | *Workshops* |
| Description | Workshops are tailored for particular groups. SD competency: Proficient. |
| Example | *"How to build…" (Wheat)*: A workshop for policymakers with the aim of creating a practical economics model. Duration: several weeks. Participants have economics and policy backgrounds. Participants develop knowledge of system dynamics modeling and software. |

*Note:* System dynamics skills improve with each level.

the analytical frameworks used (Becker 2000; Mankiw 2020). Fourth, the benefits of new teaching techniques are not always certain (Allgood et al. 2015). Consequently, factors such as the teaching load, class sizes, faculty background and their teaching experience can significantly influence the methods and materials used in undergraduate economics courses (Ahlstrom et al. 2023). Acting rationally, instructors who contemplate the adoption of new pedagogical strategies and weigh costs against benefits often favor the traditional approach (Allgood et al. 2015; Becker 2000).

## 5 | Concluding Comments

Teaching economics with system dynamics offers students structural explanations of economic behavior by emphasizing causal relationships and feedback mechanisms that govern economic systems. Students with minimal mathematical background can explore out-of-equilibrium dynamics using simulations of system dynamics models. In this paper we have outlined the authors' experiences in teaching system dynamics in a range of different economics areas and in a variety of educational programmes at undergraduate and graduate levels. Based on the literature and these experiences, we have developed a four-level course hierarchy for using system dynamics in economics teaching. We hope it can be useful for anyone who is considering introducing system dynamics in their economics instruction.

Moving forward, we suggest the following additional work:

- Conduct a comprehensive survey of economics education with system dynamics.
- Create more web-accessible teaching materials, including learning environments with corresponding instructor manuals.
- Publish case studies on using system dynamics for economics instruction that describe how to overcome potential barriers.
- Conduct empirical studies that assess the impact of system dynamics on enhancing students' learning of economic topics.

### Acknowledgments

We thank participants in the Roundtable Discussion during the 2022 International System Dynamics Conference in Frankfurt, Germany and the reviewers of our paper presented at the 2024 International System Dynamics Conference in Bergen, Norway for valuable comments and suggestions. We are also grateful to participants in the following conferences for their feedback: 50th Annual Meeting of the Eastern Economic Association, Boston, MA, February 29–March 3, 2024, and the New York State Economics Association Conference, October 25–26, 2024, Rochester, NY. Also, we thank the anonymous reviewers of this journal. However, the views and opinions expressed are the authors' own. Open access publishing facilitated by Victoria University of Wellington, as part of the Wiley - Victoria University of Wellington agreement via the Council of Australian University Librarians.

### Data Availability Statement

Data sharing not applicable to this article as no datasets were generated or analyzed during the current study.

## Appendix A

This appendix includes equations for the model in Figure 4. Each equation is followed by variable units shown in square brackets. This model was created in the modeling software Vensim PLE that is available free of charge for educational use from the company named Ventana (https://www.vensim.com). The model file is included as a supplementary attachment. These equations can be used to create the same model in the other modeling software, such as Stella Architect from *isee systems* or the online software InsightMaker, also available for free.

Perceived Price for Supply = SMOOTH (Price, Time for Producers to React to Price Changes) [$/unit]

Quantity Supplied = Supply Schedule (Perceived Price for Supply) [unit/day]

Perceived Price for Demand = SMOOTH (Price, Time for Consumers to React to Price Changes) [$/unit]

Quantity Demanded = Demand Schedule (Perceived Price for Demand) + Shift in Demand [unit/day]

Price = INTEG (Price Change, 25) [$/unit]

Price Change = ((1 − Supply-Demand Ratio) * Price)/Time to Adjust Price [$/unit/day]

Supply-Demand Ratio = Quantity Supplied/Quantity Demanded [dimensionless]

Shift in Demand = STEP (10, 10) [unit/day]

Demand Schedule ([(0, 0)-(50, 100)], (0, 100), (50, 0)) [unit/day]

Supply Schedule ([(0, 0)-(50, 100)], (0, 0), (50, 100)) [unit/day]

Time for Producers to React to Price Changes = 5 [day]

Time for Consumers to React to Price Changes = 2 [day]

Time to Adjust Price = 1 [day]

The SMOOTH function is used in system dynamics to represent expectations that are calculated as time averages. In Vensim PLE, the function SMOOTH($x$, $\tau$) has the following formulation:

$$x' = \int \frac{x - x'}{\tau} dt + x_0$$

The demand and supply curves used in the model are shown in Figure A1.

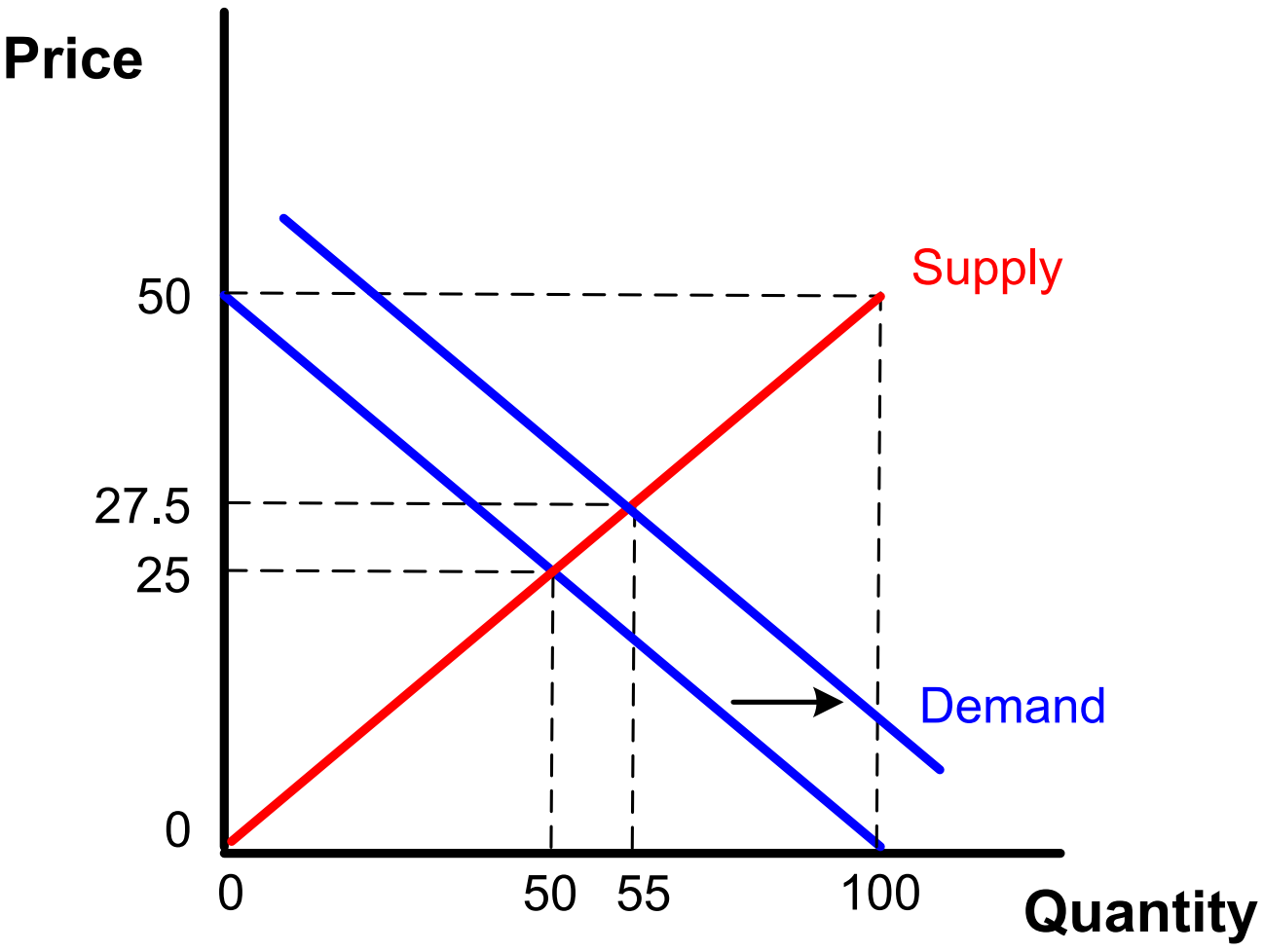

**FIGURE A1** | The demand and supply curves used in the model. The simulation starts in the initial equilibrium point, $P_1 = 25$ \$/unit and $Q_1 = 50$ unit/day. After the demand shift, the new equilibrium point is $P_2 = 27.5$ \$/unit and $Q_2 = 55$ unit/day.